\documentclass[fleqn,usenatbib]{mnras}

\usepackage{newtxtext,newtxmath}

\usepackage[T1]{fontenc}

\DeclareRobustCommand{\VAN}[3]{#2}
\let\VANthebibliography\thebibliography
\def\thebibliography{\DeclareRobustCommand{\VAN}[3]{##3}\VANthebibliography}

\usepackage{graphicx}	
\usepackage{amsmath}	
\usepackage{bm}
\usepackage{float}
\usepackage{stfloats}

\title[Transfer Learning for Transients]{Transfer Learning for Transient Classification: From Simulations to Real Data and ZTF to LSST}

\author[R. Gupta et al.]{
Rithwik Gupta,$^{1,2}$
Daniel Muthukrishna,$^{1}$\thanks{E-mail: \href{mailto:danmuth@mit.edu}{danmuth@mit.edu}}
Nabeel Rehemtulla,$^{3, 4, 5}$
Ved Shah$^{3, 4, 5}$
\\
$^{1}$Kavli Institute for Astrophysics and Space Research, Massachusetts Institute of Technology, Cambridge, MA 02139, USA\\
$^{2}$Irvington High School, 41800 Blacow Rd, Fremont, CA 94538, USA\\
$^{3}$Department of Physics and Astronomy, Northwestern University, 2145 Sheridan Road, Evanston, IL 60208, USA\\
$^{4}$Center for Interdisciplinary Exploration and Research in Astrophysics (CIERA), 1800 Sherman Ave., Evanston, IL 60201, USA\\
$^{5}$NSF-Simons AI Institute for the Sky (SkAI), 172 E. Chestnut St., Chicago, IL 60611, USA\\
}
\date{Accepted XXX. Received YYY; in original form ZZZ}

\pubyear{\the\year{}}

\begin{document}
\label{firstpage}
\pagerange{\pageref{firstpage}--\pageref{lastpage}}
\maketitle

\begin{abstract}
Machine learning has become essential for automated classification of astronomical transients, but current approaches face significant limitations: classifiers trained on simulations struggle with real data, models developed for one survey cannot be easily applied to another, and new surveys require prohibitively large amounts of labelled training data. These challenges are particularly pressing as we approach the era of the Vera C. Rubin Observatory's Legacy Survey of Space and Time (LSST), where existing classification models will need to be retrained using LSST observations. We demonstrate that transfer learning can overcome these challenges by repurposing existing models trained on either simulations or data from other surveys. Starting with a model trained on simulated Zwicky Transient Facility (ZTF) light curves, we show that transfer learning reduces the amount of labelled real ZTF transients needed by 95\% while maintaining equivalent performance to models trained from scratch. Similarly, when adapting ZTF models for LSST simulations, transfer learning achieves 94\% of the baseline performance while requiring only 30\% of the training data. These findings have significant implications for the early operations of LSST, suggesting that reliable automated classification will be possible soon after the survey begins, rather than waiting months or years to accumulate sufficient training data. 
\end{abstract}




\section{Introduction}

With the development of advanced survey telescopes, we are entering a new era for astronomical study. The Vera Rubin Observatory's Legacy Survey of Space and Time \citep[LSST;][]{Ivezic2019LSST:Products} is expected to generate tens of millions of transient alerts per night, an order of magnitude more than current state-of-the-art surveys. This unprecedented volume of data has motivated the development of numerous machine learning-based alert and light-curve classification methods \citep[e.g.][]{Charnock2016, Muthukrishna19RAPID, SupernnoovaMoller2019, Boone2019Avocado, Gomez_2020, Villar2019SuperRAENN, Villar2020SuperRAENN, Carrasco_Davis_2021, Boone_2021, Qu_2021, GHOST, Muthukrishna2022, Pimentel2023, Gomez_2023, sheng2024neuralenginediscoveringluminous, desoto2024superphotrealtimefittingclassification, Rehemtulla_2024, Shah2025}.

Past studies have been limited by three key constraints. First, classification pipelines are often trained and validated on simulated data, but struggle to generalise to real observations due to subtle differences between simulated and observed light curves \citep[e.g.,][]{Muthukrishna19RAPID}. Second, models developed for specific surveys typically require complete retraining when applied to data from different telescopes, despite the underlying physics of the transients remaining unchanged. Third, achieving high classification performance requires large amounts of labelled training data - a resource-intensive process requiring considerable time and expensive spectroscopic follow-up observations.

To address these limitations, we explore using transfer learning, a method that has been successfully applied across many fields in machine learning to adapt models trained for one task to similar applications \citep{transfer}. The technique works by reusing features and relationships that a model has already learned, even when the specific data distributions differ. This is particularly effective in two scenarios that directly address the challenges in astronomical classification: when labelled data in the target domain are limited or when training new models from scratch is computationally expensive. By reusing learned features and fine-tuning only specific components of the model, transfer learning can dramatically reduce both the amount of labelled data and computational resources needed to achieve high performance on new tasks.

The potential of transfer learning and domain adaptation is particularly relevant as we prepare for LSST. It provides a solution to the gap between simulated and real data. Deep learning models require large amounts of training data, which has led to the development of sophisticated light curve simulations \citep[e.g.,][]{KesslerPlasticcModels,elasticc} and numerous classifiers trained on these simulations \citep[e.g.,][]{Hlozek2020PLAsTiCCResults}. However, simulations do not perfectly capture real observations, making training on labelled survey data necessary \citep[e.g.,][]{Rehemtulla_2024}. While previous work has treated simulated and real data independently, with models often struggling to generalise between the two, transfer learning provides a framework to leverage features learned from simulations while adapting to real observations. Similarly, while most classifiers are built for specific surveys, transfer learning enables models trained on existing facilities to be adapted for new ones, significantly reducing the data requirements for achieving high performance. 
However, the initial model architecture determines the feasibility of transfer learning across different surveys, and not all existing classification models are suitable candidates for this approach. In this work, we incorporate the observation wavelength directly into the model to facilitate transfer learning across different surveys with varying passbands.

We demonstrate three key benefits of transfer learning for astronomical transient classification:
\begin{enumerate}
\item Transfer learning reduces the amount of data needed to generalise from simulated data to real observations, \item Transfer learning reduces the amount of labelled data needed to generalise from one survey to another, and
\item Transfer learning reduces training time for classification models.
\end{enumerate}

The paper is organised as follows. We introduce the datasets used for our experiments in Section \ref{sec:data}. We outline our experimental setup and transfer learning methodology in Section \ref{sec:methods}. We then examine the benefits of using transfer learning in Section \ref{sec:results}, demonstrating both the reduction in required training data and the ability to transfer between simulated and real observations. Finally, we discuss the broader implications of this work and future directions in Section \ref{sec:conclusion}.

\section{Data}
\label{sec:data}

In this work, we use three distinct datasets to evaluate transfer learning for transient classification: simulated light curves that match the observing properties of the Zwicky Transient Facility \citep[ZTF;][]{Bellm+2019a}, real observations from the ZTF Bright Transient Survey \citep[BTS;][]{Fremling+2020, Perley+2020, Rehemtulla_2024}, and simulated light curves matching the expected LSST observing properties. Each dataset provides different opportunities to test transfer learning between simulated and real data, as well as between different surveys. The class distribution for each dataset is summarised in Table \ref{table:datadesc}.

\begin{table*}
\begin{center}
\caption{Class distribution for the datasets used in this work.}
\label{table:ztfdistribution} 
\begin{tabular}{|c|c|c|c|c|c|c|c|c|c|c|c|c|}
\hline
Class & SNIa & SNIa-91bg & SNIax & SNIb & SNIc & SNIc-BL & SNII & SNIIn & SNIIb & TDE & SLSN-I & AGN \\
\hline
ZTF Sims & 11587 & 13000 & 13000 & 5267 & 1583 & 1423 & 13000 & 13000 & 12323 & 11354 & 12880 & 10561 \\
LSST Sims & 51745 & 12272 & 12012 & 37261 & 24513 & 10331 & 71806 & 34016 & 23408 & 28285 & 28322 & 32681 \\
BTS Data & \multicolumn{3}{c|}{|---------- 4454 (SNIa) ----------|} & \multicolumn{3}{c|}{|-------- 259 (SNIb/c) --------|} & \multicolumn{3}{c|}{|-------- 1212 (SNII) --------|} & 22 & 46 & 0 \\
\hline
\end{tabular}
\label{table:datadesc}

\end{center}
\end{table*}

\subsection{Simulated ZTF Data (ZTF Sims)}

The simulated ZTF dataset is described in \S2 of both \citet{gupta2024} and \citet{Muthukrishna2022}, and is based on the models developed for the Photometric LSST Astronomical Time-Series Classification \citep[PLAsTiCC;][]{KesslerPlasticcModels}. Each transient has flux and flux error measurements in the $g$ and $r$ passbands with a median cadence of 3 days in each band\footnote{The public MSIP ZTF survey has since changed to a 2-day median cadence.}. The dataset includes light curves from 12 transient classes: SNIa, SNIa-91bg, SNIax, SNIb, SNIc, SNIc-BL, SNII, SNIIn, SNIIb, TDE, SLSN-I, and AGN.

\subsection{ELAsTiCC (LSST Sims)}

The LSST simulations are drawn from the Extended LSST Astronomical Time Series Classification Challenge \citep[ELAsTiCC;][]{elasticc}, which builds upon the original PLAsTiCC simulations with improved models and more realistic survey properties. Light curves are simulated in all six LSST passbands ($ugrizy$) with the expected LSST cadence and depth. ELAsTiCC provides realistic class distributions based on expected LSST observation rates\footnote{ELAsTiCC also includes upsampled rare classes such as kilonovae, but we do not use these in this work.}. We use the same 12 transient classes as our ZTF simulations, with class distributions shown in Table \ref{table:datadesc} reflecting the expected rates for LSST observations.

\subsection{Bright Transient Survey Data (BTS Data)}

For real observations, we use data from the ZTF BTS. The aim of BTS is to systematically classify every bright ($m < 18.5$ mag) extragalactic transient discovered in the public ZTF alert stream \citep{Masci+2019}, providing a spectroscopically complete sample. The adopted dataset is queried and processed following the procedures in \cite{Rehemtulla_2024} and considers data up to August 2024. Light curves are available in the ZTF $g$ and $r$ filters, and we use data from three broad transient classes: SNIa, SNII, and SNIb/c, and two rarer transient classes, TDE and SLSN-I. We use these broader classifications because of the small number of transients classified in most subtypes.

\section{Methods}
\label{sec:methods}

\subsection{Classifiers}

We build a Recurrent Neural Network (RNN) classifier with Gated Recurrent Units \citep[GRU;][]{GRU} following the architecture used in \citet{Muthukrishna19RAPID} and updated in \citet{Gupta2024-icml,gupta2024}. As illustrated in Figure \ref{fig:architecture}, we include a parallel fully-connected (dense) neural network to process any available metadata, which in this work consists only of Milky Way extinction, and we do not use redshift or any host galaxy information. Contextual information such as host galaxy properties have been shown to be very useful for transient classification \citep{Foley2013ClassifyingData,GHOST} and while we do not use host information in our work, our classification framework is designed to incorporate both time-dependent and time-independent features.

While we present results using this architecture, we expect transfer learning results to be broadly consistent across other classification architectures that use flexible input formats capable of handling data from different surveys.

To enable transfer learning in this work, we use a survey-agnostic input representation. The input vector, $\bm{X}_s$, for a transient $s$, is a $4 \times N_t$ matrix, where $N_t$ is the maximum number of timesteps for any light curve in our datasets. $N_t = 656$ in this work, however most light curves have far fewer observations. Each row $i$ of the input matrix $\bm{X}_s$ contains the following information:
\begin{equation}
    \bm{X}_{si} = [\lambda_p, t_{si}, f_{si}, \epsilon_{si}],
    \label{eq:input_vector}
\end{equation}
where $t_{si}$ is the time of the $i$th observation scaled between 0 and 1, $\lambda_p$ is the median wavelength of the passband, $f_{si}$ is the flux scaled by dividing by 500, and $\epsilon_{si}$ is the corresponding flux error with the same scaling. We select a constant scaling factor of 500 as it is close to the mean flux in our dataset and enables real-time classification by avoiding peak-based scaling, which would depend on observations that may not yet be available in real-time scenarios. Our model does not require redshift information, making it useful in the early observations of LSST when redshifts may not be known for many sources.

Using this input method has two key advantages. First, it circumvents the need for interpolation between observations, which can be problematic for sparsely sampled light curves. Second, it passes all passband information through the same channel, which allows us to easily repurpose a trained model for a different survey with a different number of passbands.

To counteract class imbalances in our dataset, we use a weighted categorical cross-entropy loss to train our model. The weight of a training object $w_c$ is inversely proportional to the fraction of transients from the class $c$ in the training set,
\begin{equation}
    w_c = \frac{T}{T_c},
\end{equation}
where $T$ is the total size of the dataset and $T_c$ is the number of transients from class $c$ in the dataset. We train our models using the \texttt{Adam} optimiser \citep{Kingma+2014} over $100$ epochs using \texttt{EarlyStopping} implemented in \texttt{keras} \citep{keras}.

\subsection{Transfer Learning}

\begin{figure}
    \centering
    \includegraphics[width=\linewidth]{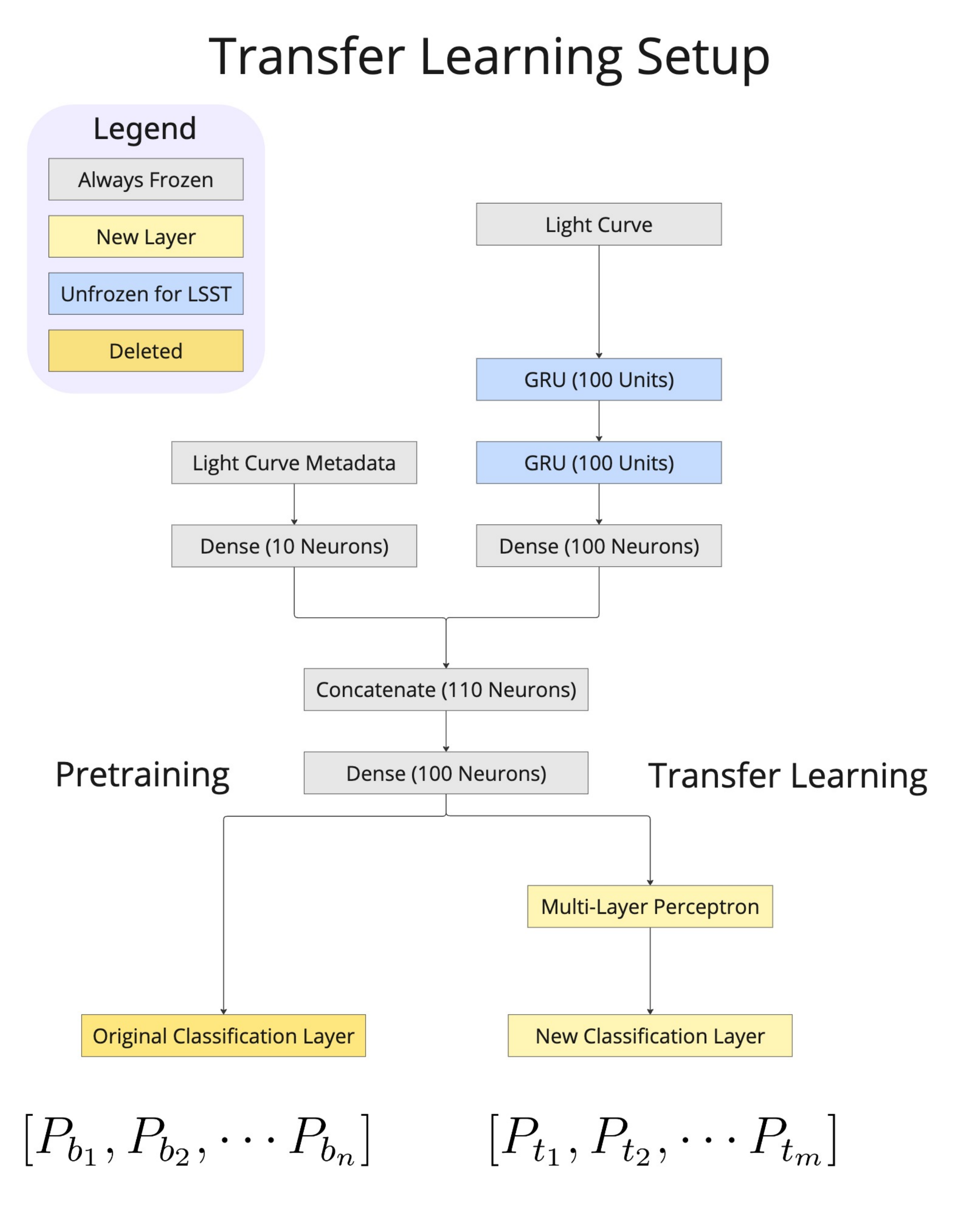}
    \caption{Schematic illustrating our neural network architecture and transfer learning methodology. The model is first pretrained on ZTF simulations to output probabilities $[P_{b_1}, P_{b_2}, ..., P_{b_n}]$ for each of the $n$ classes in the source dataset. For transfer learning, we replace the output classification layer with a new one that outputs probabilities $[P_{t_1}, P_{t_2}, ..., P_{t_m}]$ for the $m$ classes in the target dataset. When transferring to BTS data, we freeze the pretrained model and only train the new classification layer. When transferring to LSST, we also unfreeze the initial GRU layers to learn the additional passbands while keeping the dense layers frozen. The complexity of the additional Multi-Layer Perceptron (MLP) layers depends on the similarity between the source and target datasets - zero neurons for BTS (which is similar to ZTF sims) and two layers for LSST (which has different passbands). 
    }
    \label{fig:architecture}
\end{figure}

To assess the effectiveness of transfer learning, we define a source dataset and a target dataset. We first train a classifier on the source dataset, and then adapt it for the target dataset with the specific focus of analysing how much labelled data we need from the target dataset to achieve good performance. For all analyses in this work, ZTF Sims serves as the source dataset.

We are particularly interested in transfer learning from simulated to real data (ZTF to BTS) and, in preparation for the Vera Rubin Observatory, from ZTF to LSST data. Thus, we run two experiments, using either BTS Data or LSST Sims as the target dataset. In both cases, we intentionally limit the amount of labelled data available in the target dataset.

Figure \ref{fig:architecture} summarises our model architecture and experimental setup for transfer learning. We initially train a model to classify a transient into one of the $n$ classes in the ZTF Sims. After pretraining on the ZTF Sims, we replace the classification layer with a Multi-Layer Perceptron (MLP) followed by a new classification layer. The new classification layer varies in size depending on the number of classes, $m$, in the target dataset. The complexity of the MLP is determined by how different the target dataset is from the source dataset. We describe the details of each of these experiments in the following subsections.

We note that the primary benefit of transfer learning comes from pretraining rather than from specific choices about which layers to freeze, as unfreezing the entire network yields similar performance. The choice of freezing strategy primarily acts to reduce computational costs during fine-tuning.

\subsubsection{Fine-tuning for BTS}
When fine-tuning our model on the BTS data, we replace the classification layer with a new one for the $m=5$ BTS classes and freeze all other layers of the pretrained model. Since BTS and ZTF Sims share the same observing properties, we only retrain the new output layer and freeze the rest of the model weights. This prevents the network from modifying the features it has already learned about transients from its pretraining on the ZTF Sims.

\subsubsection{Fine-tuning for LSST}

Transferring from ZTF to LSST presents an additional challenge: adapting from two passbands ($gr$) to six ($ugrizy$). While our input format (see eq. \ref{eq:input_vector}) enables the model to learn a coarse relationship between flux and wavelength, the model requires retraining to capture transient behaviour across the new passbands.

For LSST, we unfreeze both the initial GRU layers (blue in Figure \ref{fig:architecture}) and the final classification layers (yellow in Figure \ref{fig:architecture}). We found that only retraining the final layers, as we did for BTS, resulted in poor performance. The initial GRU layers, which process the time-series information, need to be retrained to effectively handle the more complex and sparser LSST light curves.

While unfreezing the entire network performs similarly to our selective freezing approach, we choose to freeze the dense layers to reduce computational costs while maintaining equivalent performance.

\section{Results}
\label{sec:results}

\begin{figure*}
    \centering
    \begin{tabular}{cc}
        \includegraphics[width=0.5\linewidth]{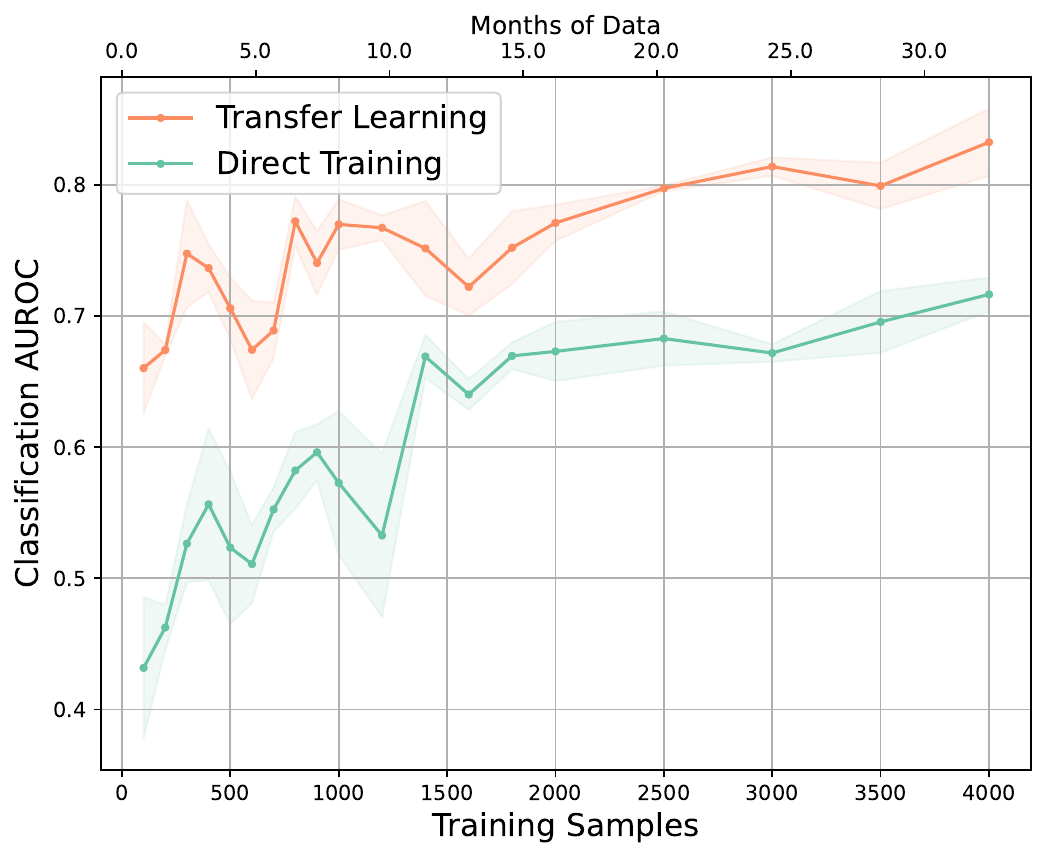}
        &
        \includegraphics[width=0.5\linewidth]{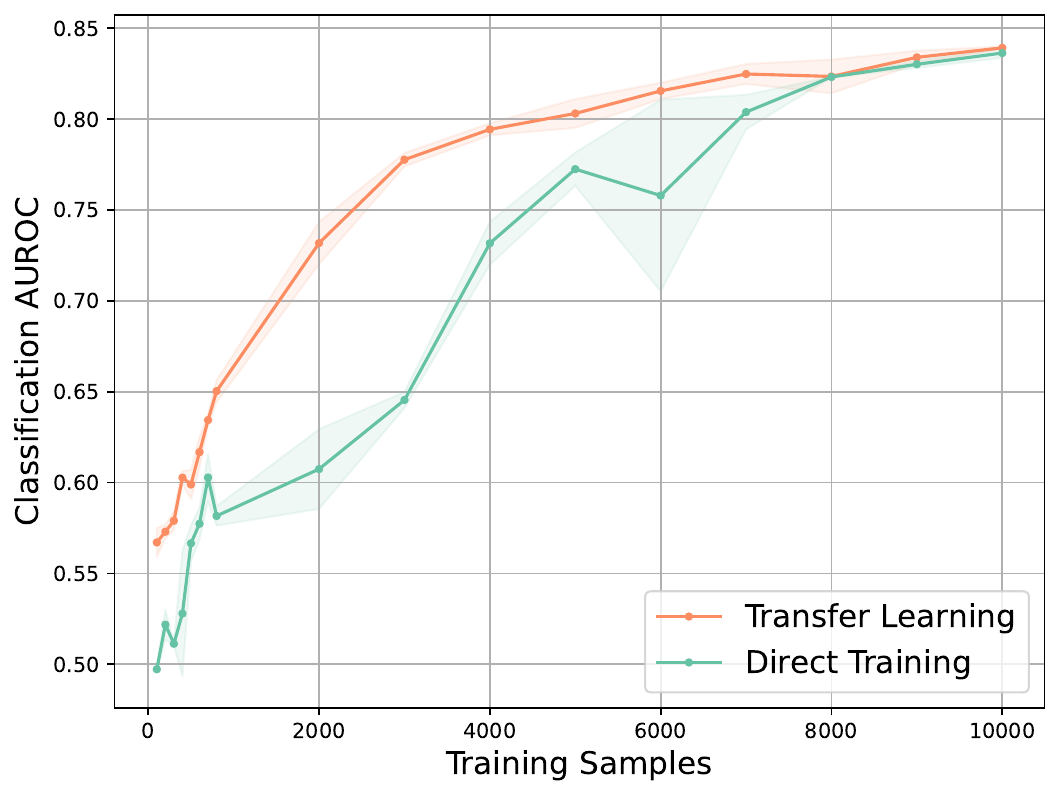}
    \end{tabular}
    \caption{Transfer learning performance from ZTF sims to BTS [left] and ZTF Sims to LSST Sims [right]. The x-axis shows the number of randomly sampled training light curves provided to the model and the y-axis shows the classification performance as the macro-averaged Area Under the Receiver Operating Characteristic (ROC) curve for each class (AUROC). The lines and shaded regions show the mean and standard deviation across ten independent training runs, respectively.}
    \label{fig:results}
\end{figure*}

To evaluate the effectiveness of transfer learning, we compare classifiers trained with transfer learning against those trained from scratch on the target dataset. We focus on two key experiments: transferring from simulated to real data (ZTF Sims to BTS) and transferring between surveys (ZTF to LSST). In both cases, we analyse how much labelled data from the target dataset is needed to achieve good classification performance. As we vary the amount of labelled data, we use 80\% of the limited data to train the model, and 20\% to validate its performance and determine optimal training time. The remaining data is used in the test set to evaluate the model.

\subsection{Simulated to Real Data}

To evaluate transfer learning from simulations to real data, we compare two approaches: fine-tuning a model pretrained on ZTF Sims and training a model directly on BTS data. For both approaches, we train with varying amounts of BTS data, sampling from the full dataset to maintain realistic class distributions. 

We evaluate the performance using the Area Under the macro-averaged Receiver Operating Characteristics Curve (AUROC). The Receiver Operating Characteristic (ROC) curve plots the False Positive Rate against the True Positive Rate over a range of classification probability thresholds.

Figure \ref{fig:results} [left] shows that transfer learning from simulated data significantly reduces the amount of real labelled data needed to build an effective classifier. A model pretrained on ZTF Sims achieves the baseline performance with just 5\% of the data. This reduction in required training data is particularly valuable given the time and expense of obtaining spectroscopic labels. We restrict our evaluation at 4,000 labelled training samples as going further would leave too few TDEs in the test set for robust evaluation. Even at this point, transfer learning provides a significant performance improvement when compared to direct training.

The performance improvement in low-data scenarios is primarily driven by minority classes, particularly TDEs and SLSNe-I, which transfer learning can effectively classify despite having very few labelled examples. These rare classes also account for the large variance observed at small training set sizes in Figure \ref{fig:results}, where classifiers become highly sensitive to each additional training sample. The class-specific results in Figure \ref{fig:bts_class} confirm that most performance variability stems from the TDE and SLSN-I classes. Importantly, transfer learning still provides substantial benefits when applied to only the common transient classes, as demonstrated in Appendix \ref{sec:majority}.

\subsection{ZTF to LSST}

To evaluate transfer learning between surveys, we compare models pretrained on ZTF Sims with those trained directly on LSST simulations. 

As depicted in Figure \ref{fig:results} [right], transfer learning significantly reduces the data requirements for developing LSST classifiers. A model pretrained on ZTF achieves 94\% of the baseline performance while requiring only 30\% of the training data. Despite the additional complexity of adapting to LSST's six passbands (compared to ZTF's two), pretraining helps the model learn features and capture some survey-agnostic patterns that are intrinsic to the astrophysical phenomena, thus reducing the amount of new training data required.
This suggests that existing classification models from current surveys such as ZTF can be efficiently adapted for LSST, enabling the rapid deployment of reliable classifiers early in the survey.

\subsection{Training Time and Computational Efficiency}
Another advantage of transfer learning beyond reducing data requirements is reduced training time, which is especially important in the era of data-driven astronomy. We find that models initialised with pretrained weights converge 25\% faster than those trained from scratch for two reasons. First, the model begins with an understanding of general transient behaviour, requiring fewer iterations to learn class-specific features. Second, with many layers frozen, there are fewer parameters to optimise during training.

The reduction in training time, while not critical on modern GPUs (training from scratch takes approximately 15 minutes on a Tesla V100), becomes significant when processing large datasets or when computational resources are limited. For example, when testing multiple model architectures or performing cross-validation with many training runs, the 25\% reduction in training time per model can substantially reduce total computation time.

\section{Conclusion}
\label{sec:conclusion}

Observations of the transient universe are entering a new era with the advent of wide-field surveys like the Vera Rubin Observatory. The unprecedented volume of data from LSST will require robust automated classification systems that can be deployed rapidly after the survey begins. Traditionally, developing reliable classifiers for a new survey requires accumulating substantial amounts of labelled data over months or years of operations. In this work, we demonstrate that transfer learning provides an effective solution to this challenge.

Our results show two significant advantages of transfer learning for astronomical transient classification:

\begin{enumerate}

\item Fine-tuning existing models requires 70\% to 95\% less labelled data than training from scratch. A model pretrained on ZTF simulations achieves equivalent performance on real data while requiring only 5\% of the labelled examples. Similarly, when adapting ZTF models for LSST, transfer learning maintains 94\% of the baseline performance with just 30\% of the training data. These results suggest that classifiers trained on existing surveys or simulations can be efficiently adapted to new data sources.

\item Transfer learning reduces computational requirements, with models converging 25\% faster than those trained from scratch. This improvement in training efficiency is particularly valuable when developing and testing multiple model architectures or when computational resources are limited.
\end{enumerate}

The success of transfer learning between both simulated and real data, as well as between different surveys, demonstrates that these models learn generalisable features of astronomical transients. This has important implications for future surveys like LSST, suggesting that reliable automated classification will be possible soon after observations begin, rather than waiting to accumulate large training sets.

Based on the expected observation rates from the ELAsTiCC dataset, which simulates the first three years of LSST operations, the survey is expected to observe approximately 50,000 transient events in its first month. Since our results demonstrate that effective classifier performance requires as few as 3,000 labelled transients, we advocate for aggressive spectroscopic follow-up efforts to enable reliable automated classification within the first few weeks of LSST operations.

Several promising directions remain for future work. First, combining data from multiple existing surveys could provide more robust pretrained models that capture a wider range of transient behaviour. 
Furthermore, applying these techniques to a broader range of astronomical phenomena could extend the benefits of transfer learning beyond transient classification.

As we prepare for the era of LSST, our results demonstrate that transfer learning will be crucial for developing the next generation of astronomical classification systems, enabling rapid scientific discovery through efficient adaptation of existing models.

\section*{Acknowledgements}

Zwicky Transient Facility access for N.R. and V.S. was supported by Northwestern University and the Center for Interdisciplinary Exploration and Research in Astrophysics (CIERA). N.R. is supported by DoE award \#DE-SC0025599.

This work used Bridges-2 at Pittsburgh Supercomputing Center through allocation PHY240105 from the Advanced Cyberinfrastructure Coordination Ecosystem: Services \& Support (ACCESS) program \citep{NSF-ACCESS-Boerner2023}, which is supported by U.S. National Science Foundation grants \#2138259, \#2138286, \#2138307, \#2137603, and \#2138296.

This work made use of the \texttt{python} programming language and the following packages: \texttt{numpy} \citep{numpy}, \texttt{matplotlib} \citep{matplotlib}, \texttt{scikit-learn} \citep{scikit-learn}, \texttt{pandas} \citep{pandas}, \texttt{astropy} \citep{astropy}, \texttt{keras} \citep{keras}, and \texttt{tensorflow} \citep{tensorflow}.

\section*{Data Availability}
The code used in this work is publicly available at \url{https://github.com/Rithwik-G/transient-transfer-learning}.

The ZTF simulations were generated using models from PLAsTiCC \citep{KesslerPlasticcModels}, available at \url{https://zenodo.org/record/2612896\#.YYAz1NbMJhE}. The light curves were simulated to match ZTF observing properties using the \texttt{SNANA} software package \citep{SNANA} with ZTF survey observing logs. These simulations build upon those used in \citet{Muthukrishna19RAPID}, updated to resolve issues with core-collapse supernovae as described in \citet{Muthukrishna2021} and \citet{Gupta2024-icml,gupta2024}. The simulated data is available upon reasonable request to the corresponding author.

The ELAsTiCC simulations
are publicly available from the LSST Dark Energy Science Collaboration (DESC) on the National Energy Research Scientific Computing Center (NERSC) at \url{https://portal.nersc.gov/cfs/lsst/DESC_TD_PUBLIC/ELASTICC/}. 

The BTS source list
is publicly available on the BTS sample explorer (\url{https://sites.astro.caltech.edu/ztf/bts/bts.php}) and light curves can be obtained from ZTF alert brokers.

\bibliographystyle{mnras}

\bibliography{references}

\begin{thebibliography}{}
\makeatletter
\relax
\def\mn@urlcharsother{\let\do\@makeother \do\$\do\&\do\#\do\^\do\_\do\%\do\~}
\def\mn@doi{\begingroup\mn@urlcharsother \@ifnextchar [ {\mn@doi@} {\mn@doi@[]}}
\def\mn@doi@[#1]#2{\def\@tempa{#1}\ifx\@tempa\@empty \href {http://dx.doi.org/#2} {doi:#2}\else \href {http://dx.doi.org/#2} {#1}\fi \endgroup}
\def\mn@eprint#1#2{\mn@eprint@#1:#2::\@nil}
\def\mn@eprint@arXiv#1{\href {http://arxiv.org/abs/#1} {{\tt arXiv:#1}}}
\def\mn@eprint@dblp#1{\href {http://dblp.uni-trier.de/rec/bibtex/#1.xml} {dblp:#1}}
\def\mn@eprint@#1:#2:#3:#4\@nil{\def\@tempa {#1}\def\@tempb {#2}\def\@tempc {#3}\ifx \@tempc \@empty \let \@tempc \@tempb \let \@tempb \@tempa \fi \ifx \@tempb \@empty \def\@tempb {arXiv}\fi \@ifundefined {mn@eprint@\@tempb}{\@tempb:\@tempc}{\expandafter \expandafter \csname mn@eprint@\@tempb\endcsname \expandafter{\@tempc}}}

\bibitem[\protect\citeauthoryear{Abadi et~al.,}{Abadi et~al.}{2015}]{tensorflow}
Abadi M.,  et~al., 2015, {TensorFlow}: Large-Scale Machine Learning on Heterogeneous Systems, \url {https://www.tensorflow.org/}

\bibitem[\protect\citeauthoryear{{Astropy Collaboration} et~al.,}{{Astropy Collaboration} et~al.}{2022}]{astropy}
{Astropy Collaboration} et~al., 2022, \mn@doi [ApJ] {10.3847/1538-4357/ac7c74}, \href {https://ui.adsabs.harvard.edu/abs/2022ApJ...935..167A} {935, 167}

\bibitem[\protect\citeauthoryear{{Bellm} et~al.,}{{Bellm} et~al.}{2019}]{Bellm+2019a}
{Bellm} E.~C.,  et~al., 2019, \mn@doi [PASP] {10.1088/1538-3873/aaecbe}, \href {https://ui.adsabs.harvard.edu/abs/2019PASP..131a8002B} {131, 018002}

\bibitem[\protect\citeauthoryear{Boerner, Deems, Furlani, Knuth  \& Towns}{Boerner et~al.}{2023}]{NSF-ACCESS-Boerner2023}
Boerner T.~J.,  Deems S.,  Furlani T.~R.,  Knuth S.~L.,   Towns J.,  2023, in Practice and Experience in Advanced Research Computing 2023: Computing for the Common Good. PEARC '23.
Association for Computing Machinery, New York, NY, USA, p. 173–176, \mn@doi{10.1145/3569951.3597559}, \url {https://doi.org/10.1145/3569951.3597559}

\bibitem[\protect\citeauthoryear{{Boone}}{{Boone}}{2019}]{Boone2019Avocado}
{Boone} K.,  2019, \mn@doi [AJ] {10.3847/1538-3881/ab5182}, \href {https://ui.adsabs.harvard.edu/abs/2019AJ....158..257B} {158, 257}

\bibitem[\protect\citeauthoryear{Boone}{Boone}{2021}]{Boone_2021}
Boone K.,  2021, \mn@doi [AJ] {10.3847/1538-3881/ac2a2d}, 162, 275

\bibitem[\protect\citeauthoryear{{Cabrera-Vives} et~al.,}{{Cabrera-Vives} et~al.}{2024}]{ATAT}
{Cabrera-Vives} G.,  et~al., 2024, \mn@doi [A\&A] {10.1051/0004-6361/202449475}, \href {https://ui.adsabs.harvard.edu/abs/2024A&A...689A.289C} {689, A289}

\bibitem[\protect\citeauthoryear{Carrasco-Davis et~al.,}{Carrasco-Davis et~al.}{2021}]{Carrasco_Davis_2021}
Carrasco-Davis R.,  et~al., 2021, \mn@doi [AJ] {10.3847/1538-3881/ac0ef1}, 162, 231

\bibitem[\protect\citeauthoryear{{Charnock} \& {Moss}}{{Charnock} \& {Moss}}{2017}]{Charnock2016}
{Charnock} T.,  {Moss} A.,  2017, \mn@doi [ApJl] {10.3847/2041-8213/aa603d}, \href {http://adsabs.harvard.edu/abs/2017ApJ...837L..28C} {837, L28}

\bibitem[\protect\citeauthoryear{Cho, van Merrienboer, Gulcehre, Bahdanau, Bougares, Schwenk  \& Bengio}{Cho et~al.}{2014}]{GRU}
Cho K.,  van Merrienboer B.,  Gulcehre C.,  Bahdanau D.,  Bougares F.,  Schwenk H.,   Bengio Y.,  2014, in Proceedings of the 2014 Conference on Empirical Methods in Natural Language Processing (EMNLP). Association for Computational Linguistics, pp 1724--1734, \mn@doi{10.3115/v1/D14-1179}

\bibitem[\protect\citeauthoryear{Chollet}{Chollet}{2015}]{keras}
Chollet F.,  2015, keras, \url{https://github.com/fchollet/keras}

\bibitem[\protect\citeauthoryear{{Foley} \& {Mandel}}{{Foley} \& {Mandel}}{2013}]{Foley2013ClassifyingData}
{Foley} R.~J.,  {Mandel} K.,  2013, \mn@doi [ApJ] {10.1088/0004-637X/778/2/167}, \href {http://adsabs.harvard.edu/abs/2013ApJ...778..167F} {778, 167}

\bibitem[\protect\citeauthoryear{{Fremling} et~al.,}{{Fremling} et~al.}{2020}]{Fremling+2020}
{Fremling} C.,  et~al., 2020, \mn@doi [ApJ] {10.3847/1538-4357/ab8943}, \href {https://ui.adsabs.harvard.edu/abs/2020ApJ...895...32F} {895, 32}

\bibitem[\protect\citeauthoryear{{Gagliano}, {Narayan}, {Engel}, {Carrasco Kind}  \& {LSST Dark Energy Science Collaboration}}{{Gagliano} et~al.}{2021}]{GHOST}
{Gagliano} A.,  {Narayan} G.,  {Engel} A.,  {Carrasco Kind} M.,   {LSST Dark Energy Science Collaboration} 2021, \mn@doi [\apj] {10.3847/1538-4357/abd02b}, \href {https://ui.adsabs.harvard.edu/abs/2021ApJ...908..170G} {908, 170}

\bibitem[\protect\citeauthoryear{Gomez, Berger, Blanchard, Hosseinzadeh, Nicholl, Villar  \& Yin}{Gomez et~al.}{2020}]{Gomez_2020}
Gomez S.,  Berger E.,  Blanchard P.~K.,  Hosseinzadeh G.,  Nicholl M.,  Villar V.~A.,   Yin Y.,  2020, \mn@doi [ApJ] {10.3847/1538-4357/abbf49}, 904, 74

\bibitem[\protect\citeauthoryear{Gomez, Berger, Blanchard, Hosseinzadeh, Nicholl, Hiramatsu, Villar  \& Yin}{Gomez et~al.}{2023}]{Gomez_2023}
Gomez S.,  Berger E.,  Blanchard P.~K.,  Hosseinzadeh G.,  Nicholl M.,  Hiramatsu D.,  Villar V.~A.,   Yin Y.,  2023, \mn@doi [ApJ] {10.3847/1538-4357/acc536}, 949, 114

\bibitem[\protect\citeauthoryear{{Gupta}, {Muthukrishna}  \& {Lochner}}{{Gupta} et~al.}{2024}]{Gupta2024-icml}
{Gupta} R.,  {Muthukrishna} D.,   {Lochner} M.,  2024, \mn@doi [arXiv e-prints] {10.48550/arXiv.2408.08888}, \href {https://ui.adsabs.harvard.edu/abs/2024arXiv240808888G} {p. arXiv:2408.08888}

\bibitem[\protect\citeauthoryear{{Gupta}, {Muthukrishna}  \& {Lochner}}{{Gupta} et~al.}{2025}]{gupta2024}
{Gupta} R.,  {Muthukrishna} D.,   {Lochner} M.,  2025, \mn@doi [RAS Techniques and Instruments] {10.1093/rasti/rzae054}, \href {https://ui.adsabs.harvard.edu/abs/2025RASTI...4...54G} {4, rzae054}

\bibitem[\protect\citeauthoryear{Harris et~al.,}{Harris et~al.}{2020}]{numpy}
Harris C.~R.,  et~al., 2020, \mn@doi [Nature] {10.1038/s41586-020-2649-2}, 585, 357

\bibitem[\protect\citeauthoryear{{Hlo{\v{z}}ek} et~al.,}{{Hlo{\v{z}}ek} et~al.}{2023}]{Hlozek2020PLAsTiCCResults}
{Hlo{\v{z}}ek} R.,  et~al., 2023, \mn@doi [ApJs] {10.3847/1538-4365/accd6a}, \href {https://ui.adsabs.harvard.edu/abs/2023ApJS..267...25H} {267, 25}

\bibitem[\protect\citeauthoryear{Hunter}{Hunter}{2007}]{matplotlib}
Hunter J.~D.,  2007, \mn@doi [Computing in Science &amp; Engineering] {10.1109/MCSE.2007.55}, 9, 90

\bibitem[\protect\citeauthoryear{{Ivezi{\'c}} et~al.,}{{Ivezi{\'c}} et~al.}{2019}]{Ivezic2019LSST:Products}
{Ivezi{\'c}} {\v Z}.,  et~al., 2019, \mn@doi [ApJ] {10.3847/1538-4357/ab042c}, \href {https://ui.adsabs.harvard.edu/abs/2019ApJ...873..111I} {873, 111}

\bibitem[\protect\citeauthoryear{Kessler et~al.,}{Kessler et~al.}{2009}]{SNANA}
Kessler R.,  et~al., 2009, \mn@doi [PASP] {10.1086/605984}, 121, 1028

\bibitem[\protect\citeauthoryear{{Kessler} et~al.,}{{Kessler} et~al.}{2019}]{KesslerPlasticcModels}
{Kessler} R.,  et~al., 2019, \mn@doi [PASP] {10.1088/1538-3873/ab26f1}, \href {https://ui.adsabs.harvard.edu/abs/2019PASP..131i4501K} {131, 094501}

\bibitem[\protect\citeauthoryear{{Kingma} \& {Ba}}{{Kingma} \& {Ba}}{2014}]{Kingma+2014}
{Kingma} D.~P.,  {Ba} J.,  2014, \mn@doi [arXiv e-prints] {10.48550/arXiv.1412.6980}, \href {https://ui.adsabs.harvard.edu/abs/2014arXiv1412.6980K} {p. arXiv:1412.6980}

\bibitem[\protect\citeauthoryear{{Masci} et~al.,}{{Masci} et~al.}{2019}]{Masci+2019}
{Masci} F.~J.,  et~al., 2019, \mn@doi [PASP] {10.1088/1538-3873/aae8ac}, \href {https://ui.adsabs.harvard.edu/abs/2019PASP..131a8003M} {131, 018003}

\bibitem[\protect\citeauthoryear{McKinney}{McKinney}{2010}]{pandas}
McKinney W.,  2010, in van~der Walt S.,  Millman J.,  eds, Proceedings of the 9th Python in Science Conference. pp 51 -- 56

\bibitem[\protect\citeauthoryear{{M{\"o}ller} \& {de Boissi{\`e}re}}{{M{\"o}ller} \& {de Boissi{\`e}re}}{2020}]{SupernnoovaMoller2019}
{M{\"o}ller} A.,  {de Boissi{\`e}re} T.,  2020, \mn@doi [MNRAS] {10.1093/mnras/stz3312}, \href {https://ui.adsabs.harvard.edu/abs/2020MNRAS.491.4277M} {491, 4277}

\bibitem[\protect\citeauthoryear{Muthukrishna, Narayan, Mandel, Biswas  \& Hlo{\v{z}}ek}{Muthukrishna et~al.}{2019}]{Muthukrishna19RAPID}
Muthukrishna D.,  Narayan G.,  Mandel K.~S.,  Biswas R.,   Hlo{\v{z}}ek R.,  2019, \mn@doi [PASP] {10.1088/1538-3873/ab1609}, 131, 118002

\bibitem[\protect\citeauthoryear{{Muthukrishna}, {Mandel}, {Lochner}, {Webb}  \& {Narayan}}{{Muthukrishna} et~al.}{2021}]{Muthukrishna2021}
{Muthukrishna} D.,  {Mandel} K.~S.,  {Lochner} M.,  {Webb} S.,   {Narayan} G.,  2021, NeurIPS 2021, \href {https://ui.adsabs.harvard.edu/abs/2021arXiv211208415M} {p. arXiv:2112.08415}

\bibitem[\protect\citeauthoryear{{Muthukrishna}, {Mandel}, {Lochner}, {Webb}  \& {Narayan}}{{Muthukrishna} et~al.}{2022}]{Muthukrishna2022}
{Muthukrishna} D.,  {Mandel} K.~S.,  {Lochner} M.,  {Webb} S.,   {Narayan} G.,  2022, \mn@doi [MNRAS] {10.1093/mnras/stac2582}, \href {https://ui.adsabs.harvard.edu/abs/2022MNRAS.517..393M} {517, 393}

\bibitem[\protect\citeauthoryear{{Narayan} \& {ELAsTiCC Team}}{{Narayan} \& {ELAsTiCC Team}}{2023}]{elasticc}
{Narayan} G.,  {ELAsTiCC Team} 2023, in American Astronomical Society Meeting Abstracts. p. 117.01

\bibitem[\protect\citeauthoryear{Pedregosa et~al.,}{Pedregosa et~al.}{2011}]{scikit-learn}
Pedregosa F.,  et~al., 2011, Journal of Machine Learning Research, 12, 2825

\bibitem[\protect\citeauthoryear{{Perley} et~al.,}{{Perley} et~al.}{2020}]{Perley+2020}
{Perley} D.~A.,  et~al., 2020, \mn@doi [ApJ] {10.3847/1538-4357/abbd98}, \href {https://ui.adsabs.harvard.edu/abs/2020ApJ...904...35P} {904, 35}

\bibitem[\protect\citeauthoryear{{Pimentel}, {Est{\'e}vez}  \& {F{\"o}rster}}{{Pimentel} et~al.}{2023}]{Pimentel2023}
{Pimentel} {\'O}.,  {Est{\'e}vez} P.~A.,   {F{\"o}rster} F.,  2023, \mn@doi [AJ] {10.3847/1538-3881/ac9ab4}, \href {https://ui.adsabs.harvard.edu/abs/2023AJ....165...18P} {165, 18}

\bibitem[\protect\citeauthoryear{Qu, Sako, Möller  \& Doux}{Qu et~al.}{2021}]{Qu_2021}
Qu H.,  Sako M.,  Möller A.,   Doux C.,  2021, \mn@doi [AJ] {10.3847/1538-3881/ac0824}, 162, 67

\bibitem[\protect\citeauthoryear{Rehemtulla et~al.,}{Rehemtulla et~al.}{2024}]{Rehemtulla_2024}
Rehemtulla N.,  et~al., 2024, \mn@doi [ApJ] {10.3847/1538-4357/ad5666}, 972, 7

\bibitem[\protect\citeauthoryear{Shah, Gagliano, Malanchev, Narayan  \& Collaboration}{Shah et~al.}{2025}]{Shah2025}
Shah V.~G.,  Gagliano A.,  Malanchev K.,  Narayan G.,   Collaboration T. L. D. E.~S.,  2025, ORACLE: A Real-Time, Hierarchical, Deep-Learning Photometric Classifier for the LSST (\mn@eprint {arXiv} {2501.01496}), \url {https://arxiv.org/abs/2501.01496}

\bibitem[\protect\citeauthoryear{{Sheng} et~al.,}{{Sheng} et~al.}{2024}]{sheng2024neuralenginediscoveringluminous}
{Sheng} X.,  et~al., 2024, \mn@doi [\mnras] {10.1093/mnras/stae1253}, \href {https://ui.adsabs.harvard.edu/abs/2024MNRAS.531.2474S} {531, 2474}

\bibitem[\protect\citeauthoryear{{Villar} et~al.,}{{Villar} et~al.}{2019}]{Villar2019SuperRAENN}
{Villar} V.~A.,  et~al., 2019, \mn@doi [ApJ] {10.3847/1538-4357/ab418c}, \href {https://ui.adsabs.harvard.edu/abs/2019ApJ...884...83V} {884, 83}

\bibitem[\protect\citeauthoryear{{Villar} et~al.,}{{Villar} et~al.}{2020}]{Villar2020SuperRAENN}
{Villar} V.~A.,  et~al., 2020, \mn@doi [ApJ] {10.3847/1538-4357/abc6fd}, \href {https://ui.adsabs.harvard.edu/abs/2020ApJ...905...94V} {905, 94}

\bibitem[\protect\citeauthoryear{Zhuang, Qi, Duan, Xi, Zhu, Zhu, Xiong  \& He}{Zhuang et~al.}{2021}]{transfer}
Zhuang F.,  Qi Z.,  Duan K.,  Xi D.,  Zhu Y.,  Zhu H.,  Xiong H.,   He Q.,  2021, \mn@doi [Proceedings of the IEEE] {10.1109/JPROC.2020.3004555}, 109, 43

\bibitem[\protect\citeauthoryear{{de Soto} et~al.,}{{de Soto} et~al.}{2024}]{desoto2024superphotrealtimefittingclassification}
{de Soto} K.~M.,  et~al., 2024, \mn@doi [\apj] {10.3847/1538-4357/ad6a4f}, \href {https://ui.adsabs.harvard.edu/abs/2024ApJ...974..169D} {974, 169}

\makeatother
\end{thebibliography}

\appendix

\section{Class-based Performance}

\label{sec:class_based}
\subsection{BTS}
\begin{figure}
    \centering
    \includegraphics[width=\linewidth]{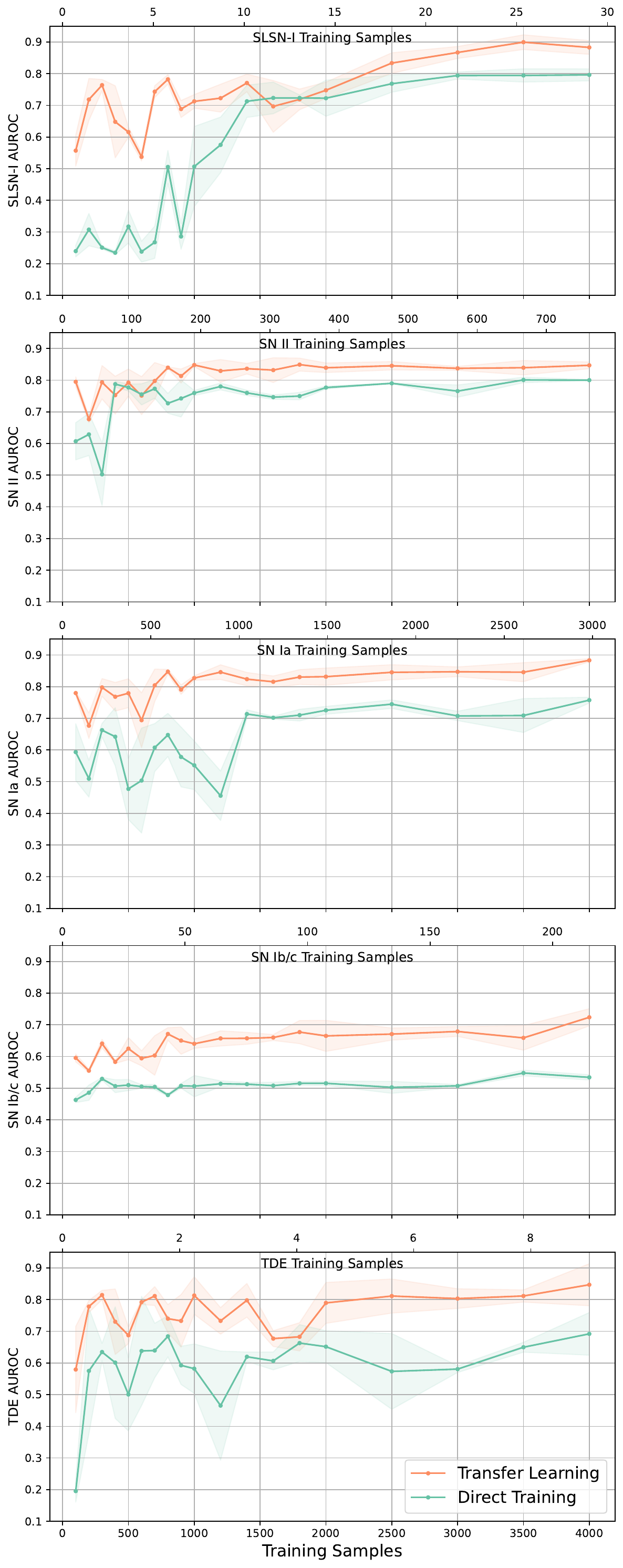}
    \caption{Class-specific classification AUROCs for transfer learning from simulated ZTF data to real BTS data.}
    \label{fig:bts_class}
\end{figure}

Figure \ref{fig:bts_class} shows the class-specific performance of transfer learning compared to direct training on BTS data. Transfer learning provides significant improvement across all transient classes, with the most pronounced benefits observed for rare classes such as SLSNe-I and TDEs, which have very few training samples.

\subsection{LSST}

\begin{figure}
    \centering
    \includegraphics[width=\linewidth]{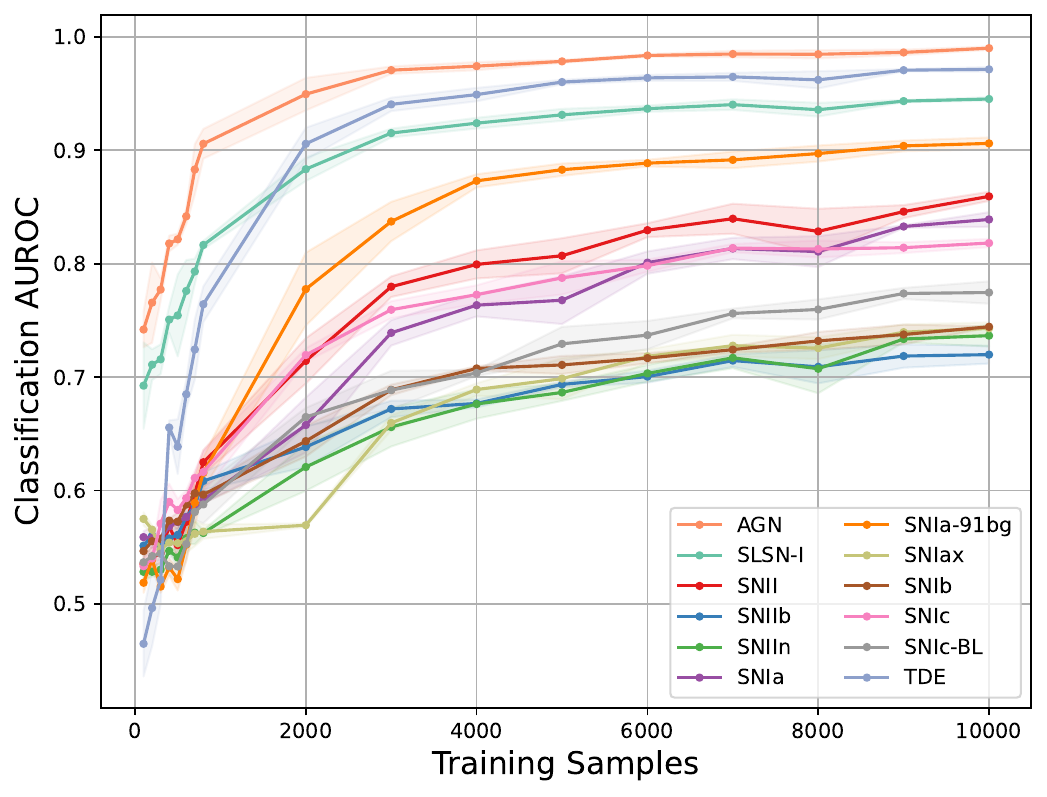}
    \caption{Class-specific classification AUROCs for transfer learning from ZTF sims to LSST sims.}
    \label{fig:lsst_class}
\end{figure}

Figure \ref{fig:lsst_class} shows the class-specific performance for transfer learning from ZTF simulations to LSST simulations. The relative performance across different transient classes reflects trends observed in other photometric classifiers such as RAPID \citep{Muthukrishna19RAPID}, ATAT \citep{ATAT}, and ORACLE \citep{Shah2025}. Classes with distinctive photometric signatures, such as AGNs with their characteristic stochastic variability over long timescales, achieve higher classification accuracy than supernova subtypes with more subtle differences. For example, distinguishing between SN-Ib and SN-Ic based solely on photometric rise times remains challenging, as these subtle features do not always manifest clearly in broad-band observations.

\section{Performance without Minority Classes}

\label{sec:majority}

\begin{figure}
    \centering
    \includegraphics[width=\linewidth]{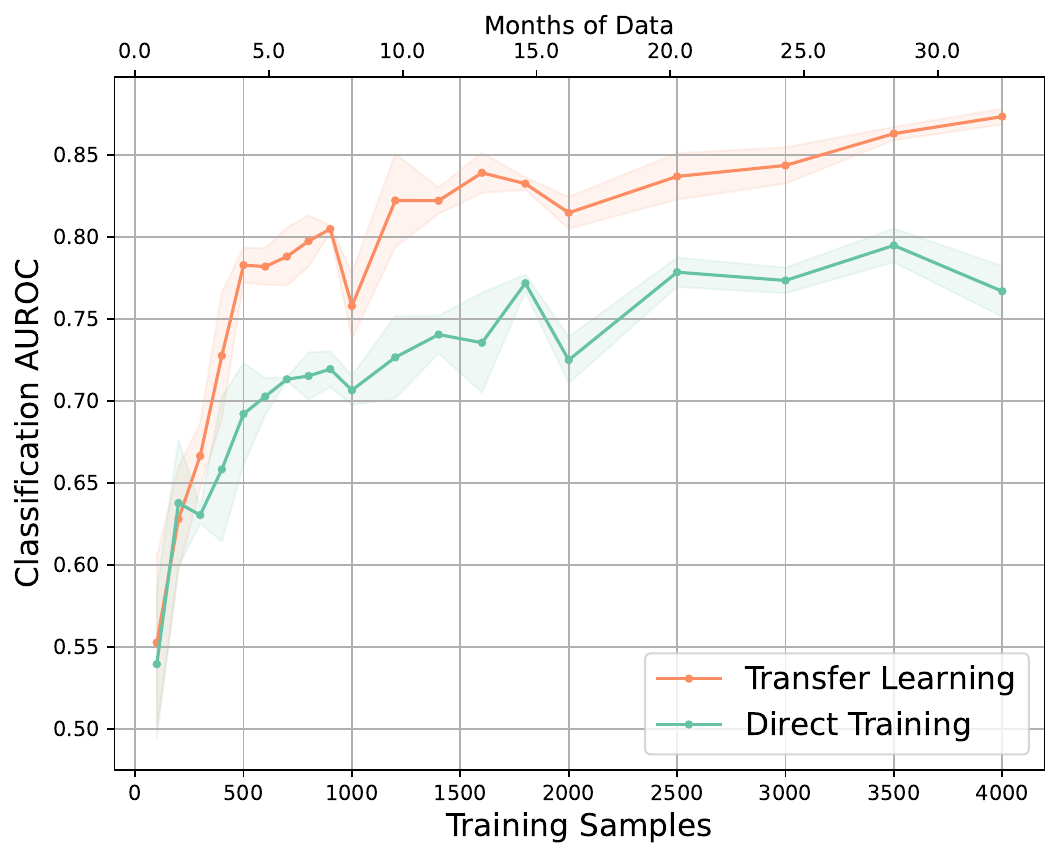}
    \caption{A recreation of Figure \ref{fig:results} [left] with TDEs and SLSNe-I omitted.}
    \label{fig:bts_three}
\end{figure}

The transfer learning performance improvement shown in Figure \ref{fig:results} [left] is partly attributed to minority classes (SLSN-I and TDE), which have very few labelled observations but are relatively easy to classify \citep{Muthukrishna19RAPID, gupta2024}. To test whether transfer learning benefits extend beyond these rare classes, Figure \ref{fig:bts_three} shows the same analysis using only the three most common transient types (SNIa, SNII, and SNIb/c). Transfer learning still provides substantial improvement even when focusing solely on these well-represented classes.

\bsp	
\label{lastpage}
\end{document}